\documentclass[a4paper,preprint,unsortedaddress,onecolumn]{revtex4-1}
\usepackage{geometry} 
\usepackage{enumerate}
\usepackage{amsmath}
\usepackage{amssymb}
\usepackage{graphicx}
\usepackage{color}
\usepackage{tabularx}
\usepackage{gensymb}
\usepackage{multirow}

\usepackage[footnotesize]{subfigure}
\usepackage{color}
\definecolor{light}{gray}{0.50}
\definecolor{heavy}{gray}{0.35}
\definecolor{black}{gray}{0.0}
\definecolor{dgreen}{rgb}{0.0,0.7,0}
\definecolor{dred}{rgb}{0.9959,0,0}
\definecolor{green}{rgb}{0.0,0.99599,0.0}
\definecolor{purple}{rgb}{0.6,0.0,0.4}

\newcommand{\red}[1]{\textcolor{black}{#1}}

\begin{document}
\title{Towards Open Boundary Molecular Dynamics Simulation of Ionic Liquids}
\author{Christian Krekeler}
\affiliation{Institute for Mathematics, Freie Universit\"at Berlin, Germany}
\email{ch.krekeler@fu-berlin.de}
\author{Luigi Delle Site}
\affiliation{Institute for Mathematics, Freie Universit\"at Berlin, Germany}
\email{luigi.dellesite@fu-berlin.de}
\begin{abstract}
We extend the use of the adaptive resolution method (AdResS) in its Grand Canonical-like version (GC-AdResS) to the molecular dynamics simulation of 1,3-dimethylimidazolium chloride. We show that the partitioning of the total system in a subsystem of interest with atomistic details and a reservoir of coarse-grained particles leads to satisfactory results. The challenging aspect of this study, compared to previous AdResS simulations, is the presence of charged particles and the necessity of addressing the question about the minimal physical input needed to model the coarse-grained particles in the reservoir. We propose two different approaches and show that in both cases they are sufficient to capture the decisive physical characteristics that allow a valid system-reservoir coupling. The technical satisfactory result paves the way for multiscale analysis of ionic liquids and for truly open boundary molecular simulations.
  
\end{abstract}

\maketitle


\section{Introduction}
Ionic liquids (IL's) are gaining great popularity due to the range of amazing properties and potential groundbreaking developments in chemistry, physics, biology and materials science \cite{wein}. The possibility of designing such systems at molecular level has led to a large effort by theoreticians; the aim is to understand them deeply in the perspective of manipulating and optimizing their chemical design for obtaining large scale properties on demand (see e.g. the themed collection of Phys. Chem. Chem. Phys., {\bf 12}, 1629-1629 (2010) and Faraday Discussions {\bf 154}, 1-484 (2012)). In this respect the field of molecular simulation has been particularly active and has led to important progress (see e.g.Ref.\cite{chemphyschemrev} and references therein) . Molecular simulation, by construction, allows for a molecular-based understanding of large scale properties and thus represents the most powerful tool to draw strategies for development and use of IL's.
In previous work \cite{krekil,jctclf,fdisc} we have shown that IL's (or at least a large class of them), are characterized by a local scale where important properties, which intuitively one may expect to be dominated by large scale correlations, are instead highly localized and depends only on the immediate neighbouring molecules. These results became the inspiration for the extension of the AdResS method \cite{jcp1,annurev} in the Grand Canonical-like version \cite{jctchan,prx,njp,pre16} to the study of IL's. In fact, AdResS allows to have full atomistic details in a region of interest and couple it with a region with very generic, coarse-grained, molecular model, without atomistic details, that acts as a reservoir of particle and energy; this implies that one could use GC-AdResS as a tool to identify the essential atomistic degrees of freedom required to have a certain property. This idea was already successfully used in previous AdResS studies to infer about the locality of solvation properties for hydrophobic molecules in water \cite{jcpcov} and about the locality of the IR spectrum in water \cite{lucpc}. We believe that it would be useful to have a similar tool of analysis also for ionic liquids, in this perspective, the aim of this work to show the technical feasibility of an AdResS study of IL's. The challenging aspect for the application of (GC-)AdResS to IL's is the presence of explicit ions and the capability of building coarse-grained procedure for a physically valid system-reservoir coupling. Previous work has already dealt with the presence of explicit ions \cite{njpmatej} but only for dilute solutions in water, that is, the main process was dominated by the solvating character of water. For ionic liquids we must make a step forward and treat, inspired by the previous work of water-salt solution, a dense liquid of positively and negatively charged ions. We will report about two different approaches to build sufficiently accurate coarse-grained models for our study; one where the coarse-grained models does not carry charges but reproduces the ion-ion radial distribution functions of a full atomistic simulation (according to the so called, Inverse Boltzmann Iteration , IBI, procedure \cite{ibi}) and one where ions in the coarse-grained model carry the proper charge (according to the atomistic model) and at the same time reproduce, as for the neutral model,  ion-ion radial distribution functions of a full atomistic simulation (IBI with explicit charge-charge interaction). In this work we have chosen  1,3-dimethylimidazolium chloride as test system of our simulation because it is complex enough to test the general robustness of our simulation method but at the same time is simple enough for allowing a large number of numerical tests in a straightforward way and without particularly expensive computational efforts. Beside the conceptual use of AdResS as a tool of analysis that identifies the essential atomistic degrees of freedom, one may want to consider the possible gain in terms of computational saving due to the drastic reduction of the number of degrees of freedom. Currently, the conceptually elaborated version of GC-AdResS is implemented in GROMACS \cite{gromacs} and its computational architecture is not yet optimized in such a code thus the computational gain occurs only for very large systems and it is still modest. \red{The current performance gain for systems of 20.000-40.000 ion pairs, for the ionic liquid considered in this work, is a speedup factor of 1.5-2.0  compared to the performance a full atomistic simulation; the precise factor depends on the size of the atomistic region and of the transition region.} However, independently of future technical optimization of the code, the success, at conceptual level, of the use of AdResS for IL's paves the way for the simulation of open systems with virtually an infinite size when \red{ the implementation of AdResS which is coupled to a continuum \cite{matejcont1,matejcont2,julich} is used}; for IL's this may certainly represent an interesting future 
perspective. 

\section{gc-AdResS approach}
\label{adress}
\begin{figure}
   \centering
   \includegraphics[width=0.75\textwidth]{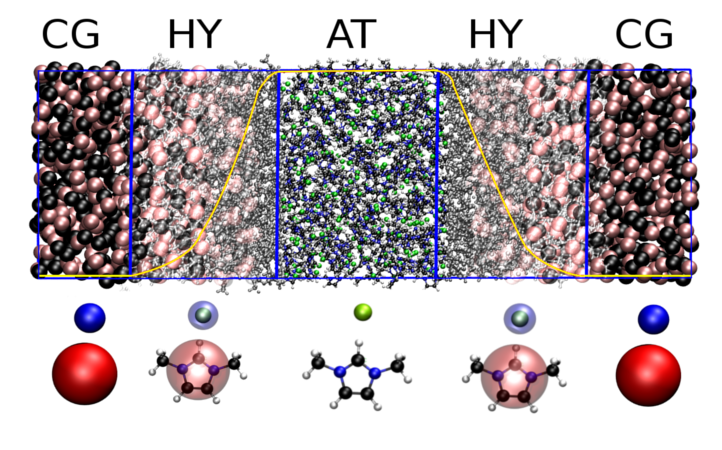}
   \caption{(Top) Pictorial representation of the GC-AdResS scheme; CG indicates the coarse-grained region, HY the hybrid region where atomistic and coarse-grained forces are interpolated via $w(x)$ (in yellow) and AT is the atomistic region (that is the region of interest). (Bottom) The molecular representation of the atomistic, hybrid and coarse-grained molecule, according to the region they belong to.}
 \label{cartoon}
 \end{figure}

The original AdResS method for Molecular Dynamics \cite{jcp1,pre1} was developed following the intuitive principle that the coupling of different regions where molecules have different molecular resolution should be done in such a way that the passage from the dynamics of one resolution (region) to another must be smooth enough so that the perturbation in each region is negligible. From the computational point of view this principle was implemented via a space dependent interpolation formula for the force acting between two molecules, $\alpha,\beta$ (see also Fig.\ref{cartoon}, for a pictorial representation):
\begin{equation}
F_{\alpha \beta} = w(X_{\alpha})w(X_{\alpha})F_{\alpha\beta}^{AT} + [1 - w(X_{\alpha})w(X_{\alpha})]F_{\alpha\beta}^{CG}
\end{equation}  
where $F_{\alpha\beta}^{AT}$ is the force derived from the atomistic potential and $F_{\alpha\beta}^{CG}$ is the force derived from the coarse-grained potential. The interpolating (switching) function is defined as:
\begin{equation*}
    w(x) = \begin{cases}
               1               & x < d_{AT} \\
               cos^{2}\left[\frac{\pi}{2(d_{\Delta})}(x-d_{AT})\right]   & d_{AT} < x < d_{AT}+d_{\Delta}\\
               0 & d_{AT} + d_{\Delta}< x
           \end{cases}
\end{equation*}
where, $d_{AT}$ is the size of the atomistic and $d_{\Delta}$ is the size of the hybrid (transition) region as reported in Fig.\ref{cartoon}; $x$ is the 
$x$-coordinate of the center of mass of the molecule.

The weighting function smoothly goes from 0 to 1 in the transition region so that the practical effect is that a coarse-grained molecule transforms into an atomistic molecule and vice versa. The system is coupled to an external thermostat that takes care of supplying or removing the required energy for the thermodynamic stability. This simple approach led to numerically highly satisfactory results for various systems \cite{pol,wat1,wat2,jcpcov,prlado,pccpado,jctckb,jcppara} and brought the need of finding a deeper conceptual justification. In fact in the next step the basic thermodynamic relations for a proper coupling were defined and then this led to the derivation of a one-particle force acting on the center of mass of the molecule in the hybrid region, named thermodynamic force, ${\bf F}_{thf}(x)$. Such force assures that the effective chemical potential of the system is that of the atomistic resolution \cite{jcpsimon}. Later on the thermodynamic force was derived in a more rigorous way, considering the atomistic and the coarse-grained region as two open regions interfaced by a filter (hybrid region) which is defined by the balance of the Grand Potential, as if each of the two region is in equilibrium with a finite particle reservoir (the other region) :$p_{AT}+\rho_{0}\int_{\Delta}{\bf F}_{thf}({\bf r})d{\bf r}=p_{CG}$,
  with $p_{AT}$  the reference pressure of the atomistic system (region), $p_{CG}$ the pressure of the coarse-grained model and $\rho_{0}$ is the reference molecular density of the atomistic system. The thermodynamic force was then written as the gradient of the particle density in an iterative form which is computationally highly convenient \cite{prl12}: $F_{k+1}^{thf}(x)=F_{k}^{thf}(x) - \frac{M_{\alpha}}{[\rho_{ref}]^2\kappa}\nabla\rho_{k}(x)$,
$M_{\alpha}$ is the mass of the molecule, $\kappa$ a constant chosen to optimize the calculation, $\rho_{k}(x)$ is the molecular density at the $k$-th iteration as a function of the position in the transition (hybrid) region.
The choice of the convergence criterion depends on the accuracy needed for the simulation, however, as a rule of thumb, $\rho_{final}-\rho_{0}$ should not be larger than $10\%$ in the transition region. Finally, in the last few years the method has found mathematical and physical rigorous formalization either in terms of a global Hamiltonian approach (H-AdResS) \cite{raff1,raff2} or in terms of Grand Ensemble approach where the coarse-grained region is composed by a liquid of simple spheres whose minimal request is that it acts as a reservoir for the atomistic region (GC-AdResS) \cite{jctchan,prx,njp,pre16}; the two starting points are of course compatible \cite{kreisepl,ralfepjst,prejinglong}. In the meanwhile the AdResS method in each of its revised current formulations has been successfully applied to a rather large class of liquids, mixtures and solvation processes \cite{kk15,kk161,kk162,mat1,mat2,mat3,luimu,lujcppi,lucpc,lupolj}. In this work we consider the GC-AdResS approach and require that the coarse-grained region is characterized by some minimal physical input so that numerical results in the atomistic region are satisfactory when compared with the results obtained in the equivalent sub-region of a full atomistic simulation. In the next section we propose two coarse-grained models and then show that in both cases results are highly satisfactory.
\section{Coarse-Grained models}
\begin{figure}
   \centering
   \includegraphics[width=0.85\textwidth]{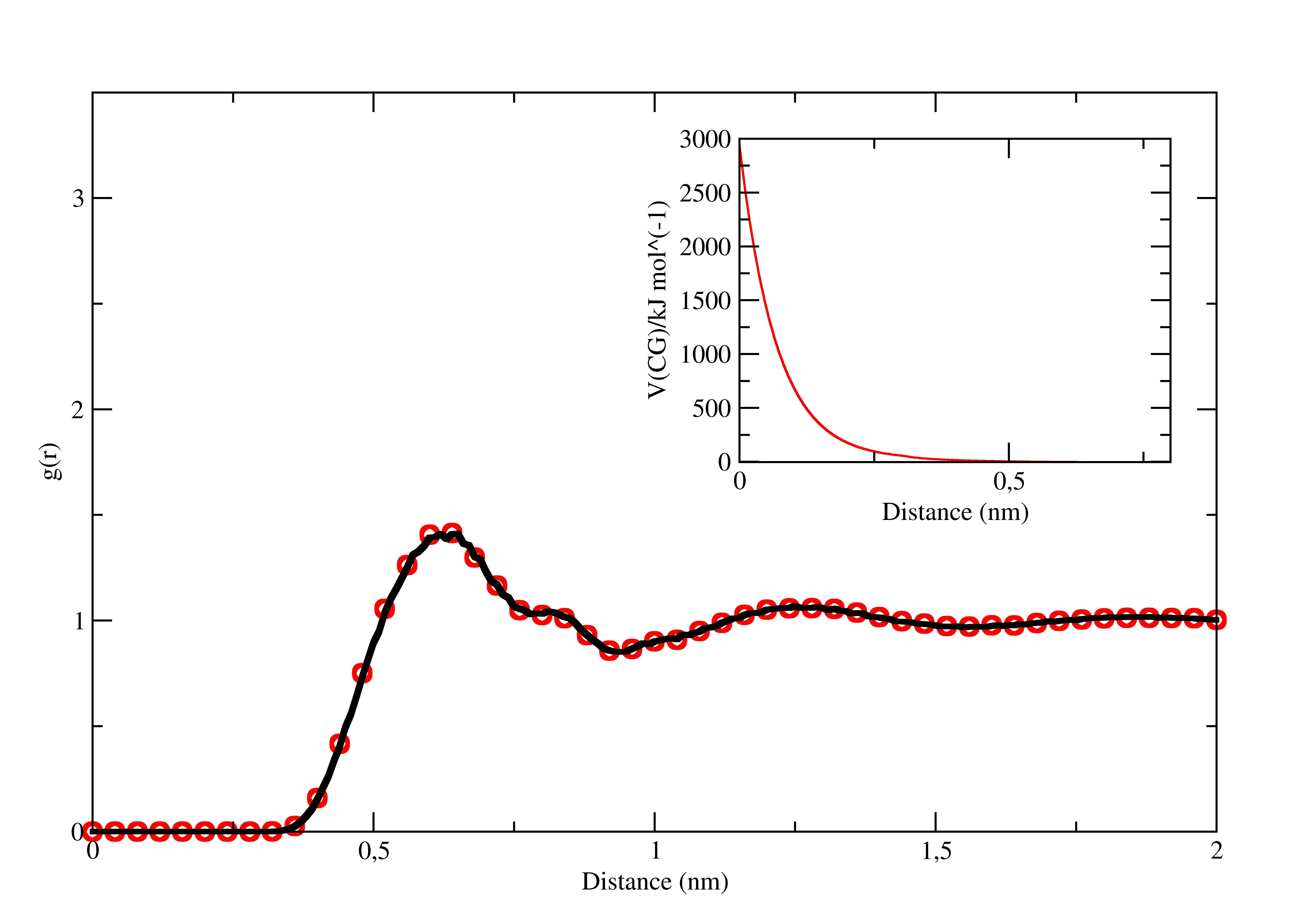}
   \caption{Main figure: Anion-anion radial distribution function from atomistic simulation (black line), compared with the anion-anion distribution function obtained from a coarse-grained simulation (red circles) which employs the IBI potential for the uncharged coarse-grained model (inset).}
 \label{anannc}
 \end{figure}

\begin{figure}
   \centering
   \includegraphics[width=0.85\textwidth]{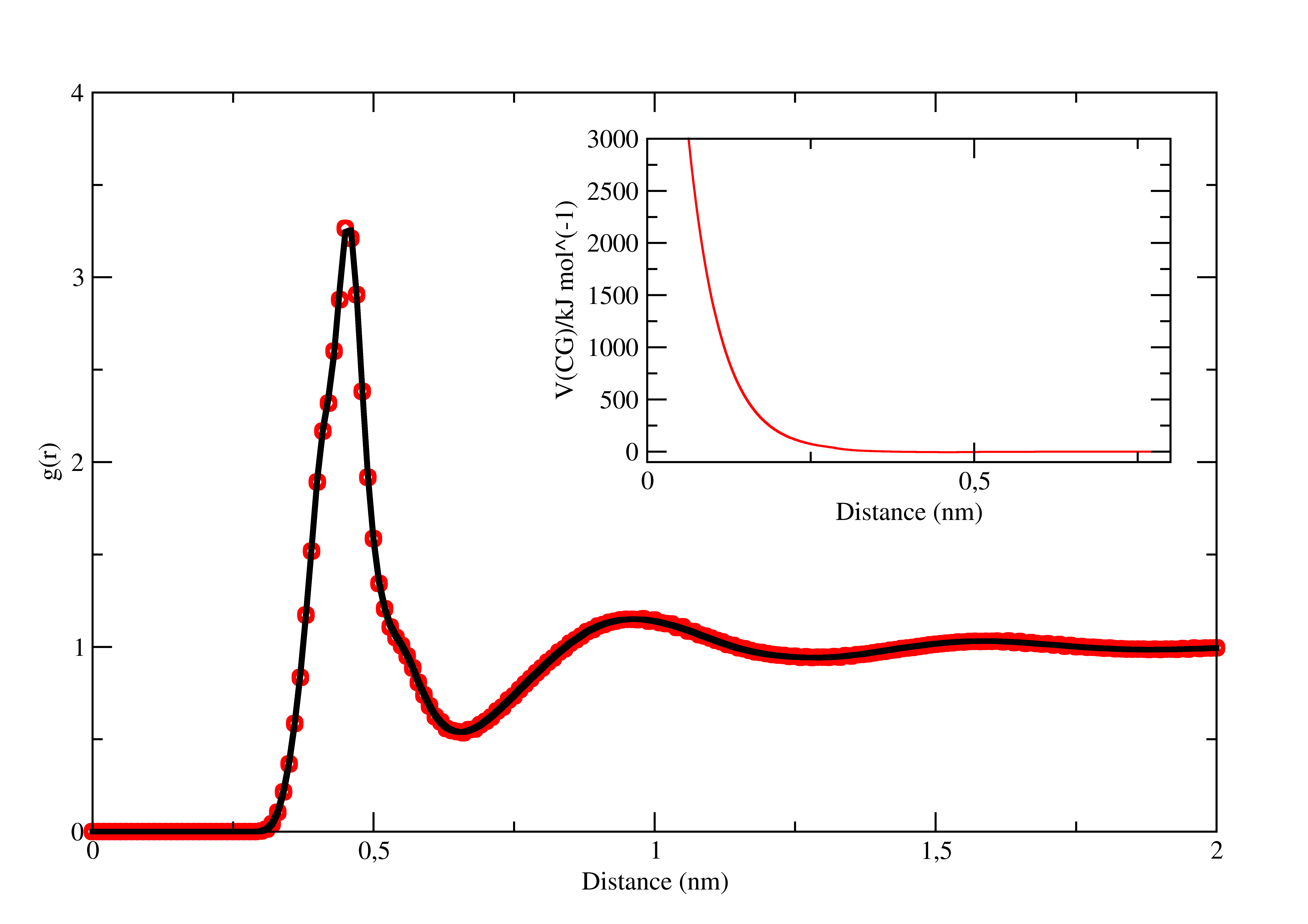}
   \caption{Equivalent of Figure \ref{anannc} for the anion-cation case.}
 \label{ancatnc}
 \end{figure}

\begin{figure}
   \centering
   \includegraphics[width=0.85\textwidth]{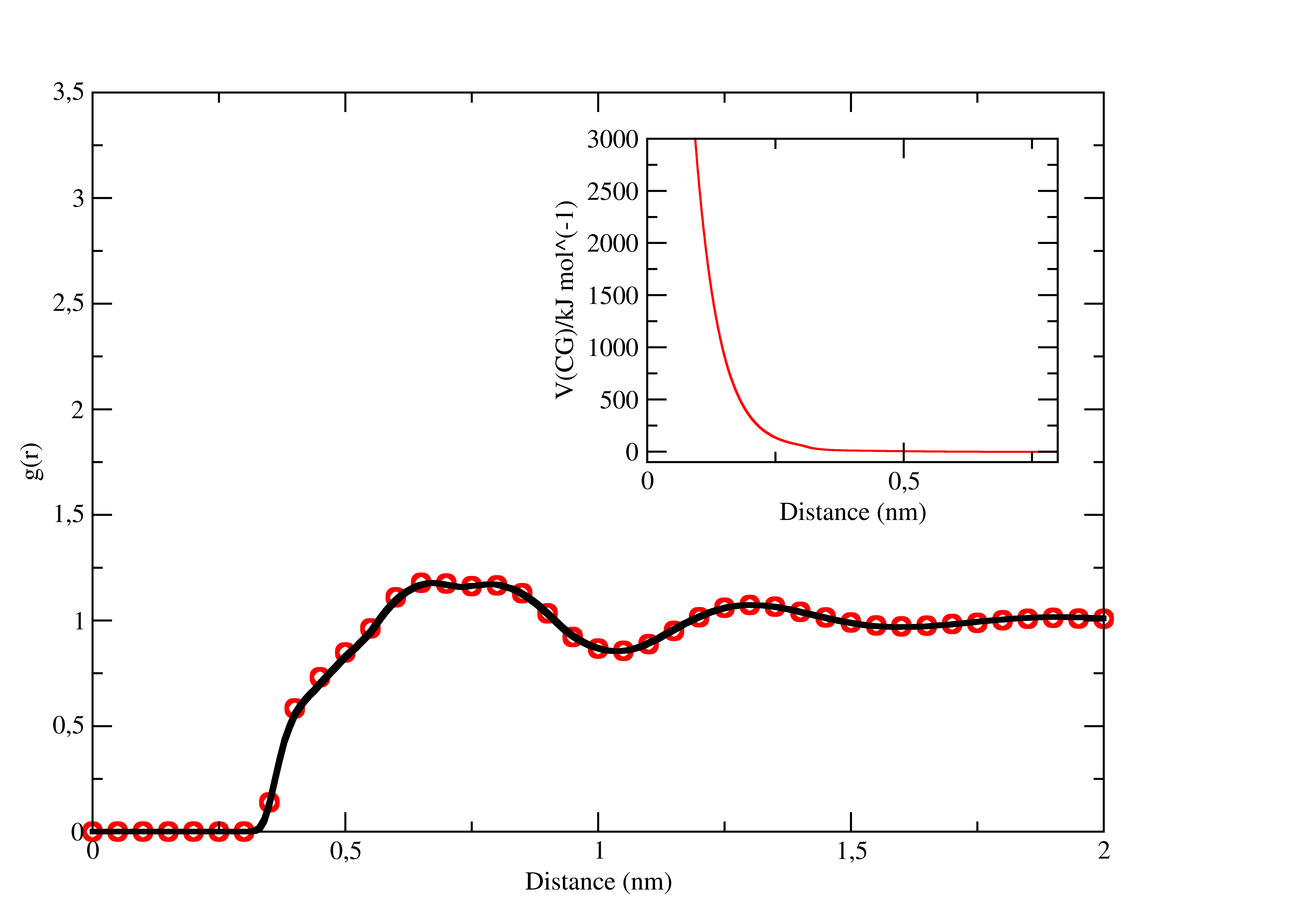}
   \caption{Equivalent of Figure \ref{anannc} for the cation-cation case.}
 \label{catcatnc}
 \end{figure}

In previous work of AdResS it has been shown that coarse-grained models as simple as liquids of generic spheres which reproduce only the particle density and the temperature of the atomistic liquid are already sufficient to act as a valid reservoir of particles and energy for the atomistic region (thus the definition of Grand Canonical-like AdResS) \cite{jctchan,prx,njp,lujcppi,lucpc}. In the group of Kremer an even more drastic step was taken and the reservoir is built using a gas of generic spheres \cite{gas}. The robustness of the AdResS is in the derivation of the thermodynamic force (from first principles) in the hybrid region; in fact it has been shown that a necessary (and numerically sufficient) condition for an accurate Grand Canonical distribution in the atomistic region is that the particle density in the hybrid region must be (ideally) equal to that of the atomistic region. Next, it has been shown that such condition and the action of the thermostat to keep the same temperature everywhere corresponds to automatically fix the chemical potential at the value one would obtain from a full atomistic simulation in the same thermodynamic conditions. The systems treated in such case where liquids and mixture of small neutral molecules. The question of how to treat the presence of ions in the system was then addressed in Ref.\cite{njpmatej}. In such a case the study concerns the solvation of Na$^{+}$ and Cl$^{-}$ in water at concentrations compatible with biological conditions. The study was aimed at extending the use of AdResS to the solvation of biological molecules and the presence of corresponding counter-ions \cite{mat3}. The work of Ref.\cite{njpmatej} represents an important starting point for the extension of the idea to liquids with high concentration of ions as for example IL's treated here. It must be clarified, and this is actually the most relevant point of our work, that we do not intend to develop a coarse-grained model of IL's that can be used for full coarse-grained simulations {\it per se}, our aim, as already underlined in the introduction, is to develop coarse-grained models which are sufficiently good to act as a reservoir for the atomistic region; such region is the only region of physical interest in our study, while the hybrid and coarse-grained region are only of technical/computational interest (i.e. to be as efficient as possible in numerical terms). Compared to the approach of Ref.\cite{njpmatej} we make a further drastic step, in the very spirit of Refs.\cite{jctchan,prx,njp} and of Ref.\cite{gas}, and consider as a first approach a coarse-grained model without explicit charges; each ion has only one interaction site, that is its center of mass (trivial for Cl$^{-}$), as pictorially illustrated in Fig.\ref{cartoon}. The interaction potentials of such a coarse-grained model is derived using the Inverse Boltzmann Iteration procedure (IBI) \cite{ibi} so that the coarse-grained potentials obtained reproduce the anion-anion, anion-cation, cation-cation full atomistic radial distribution functions respectively (See Figs\ref{anannc},\ref{ancatnc},\ref{catcatnc}). 

\begin{figure}
   \centering
   \includegraphics[width=0.85\textwidth]{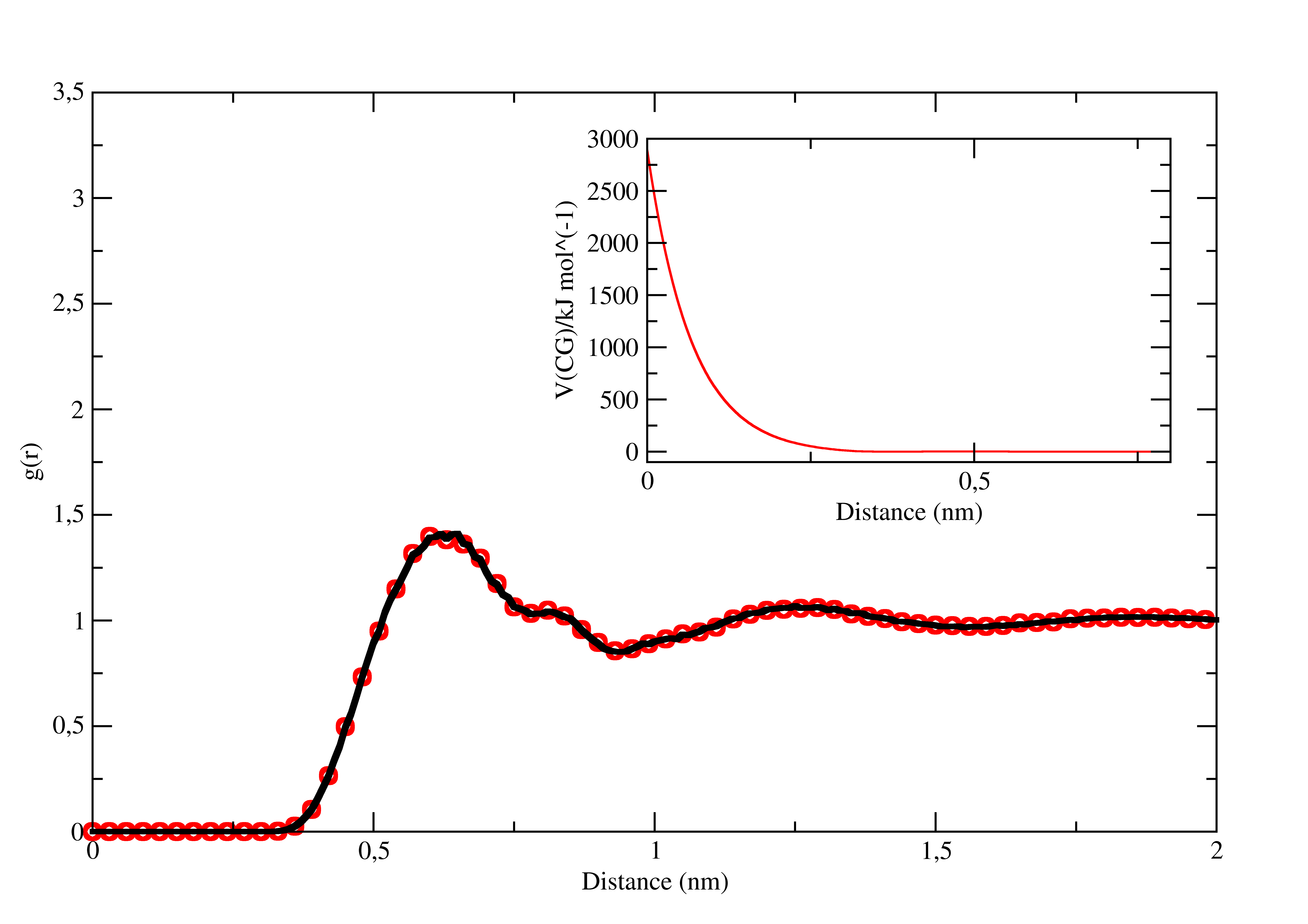}
   \caption{Equivalent of Figure \ref{anannc} for the charged coarse-grained model.}
 \label{ananc}
 \end{figure}

\begin{figure}
   \centering
   \includegraphics[width=0.85\textwidth]{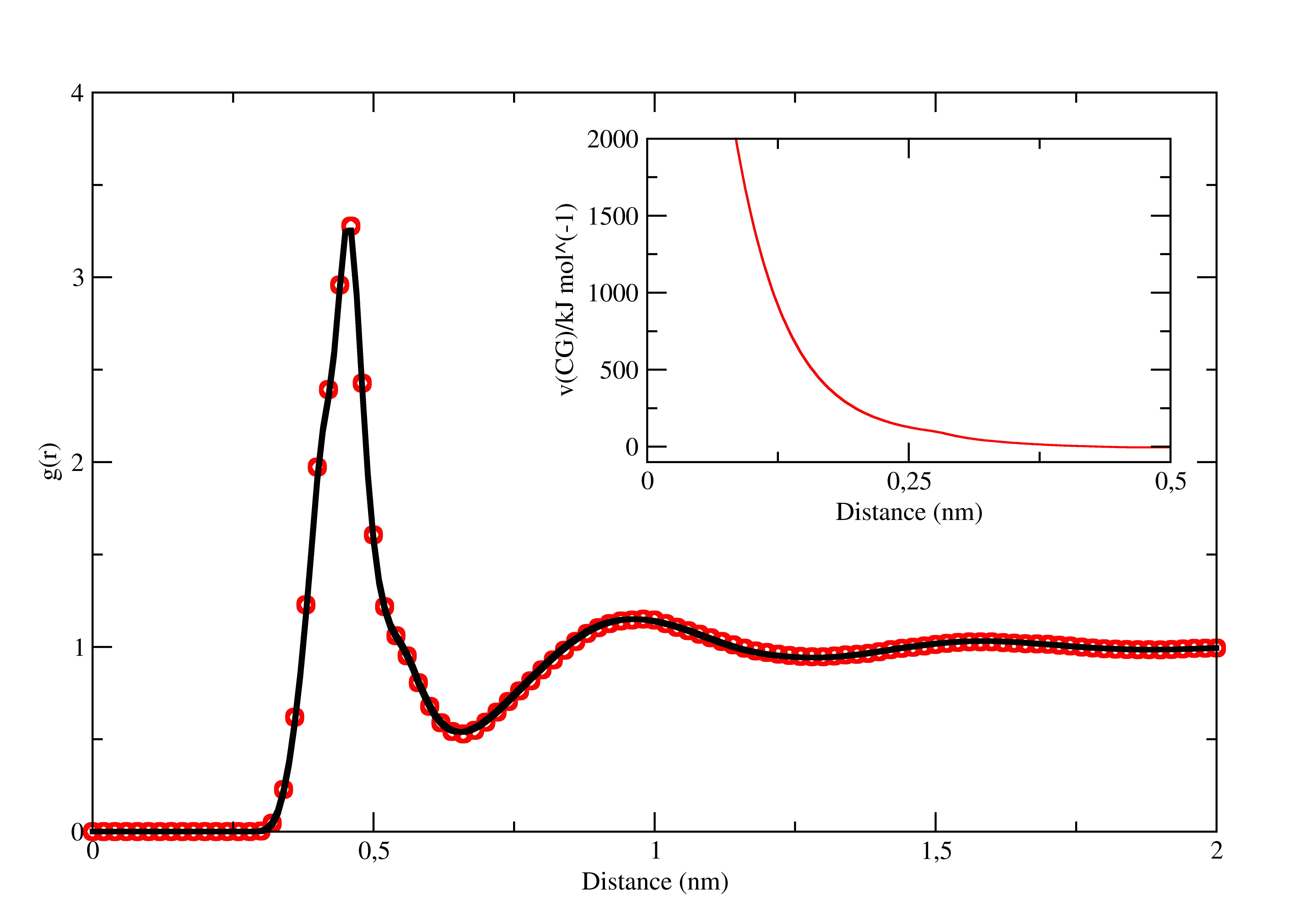}
   \caption{Equivalent of Figure \ref{ancatnc} for the charged coarse-grained model.}
 \label{ancatc}
 \end{figure}

\begin{figure}
   \centering
   \includegraphics[width=0.85\textwidth]{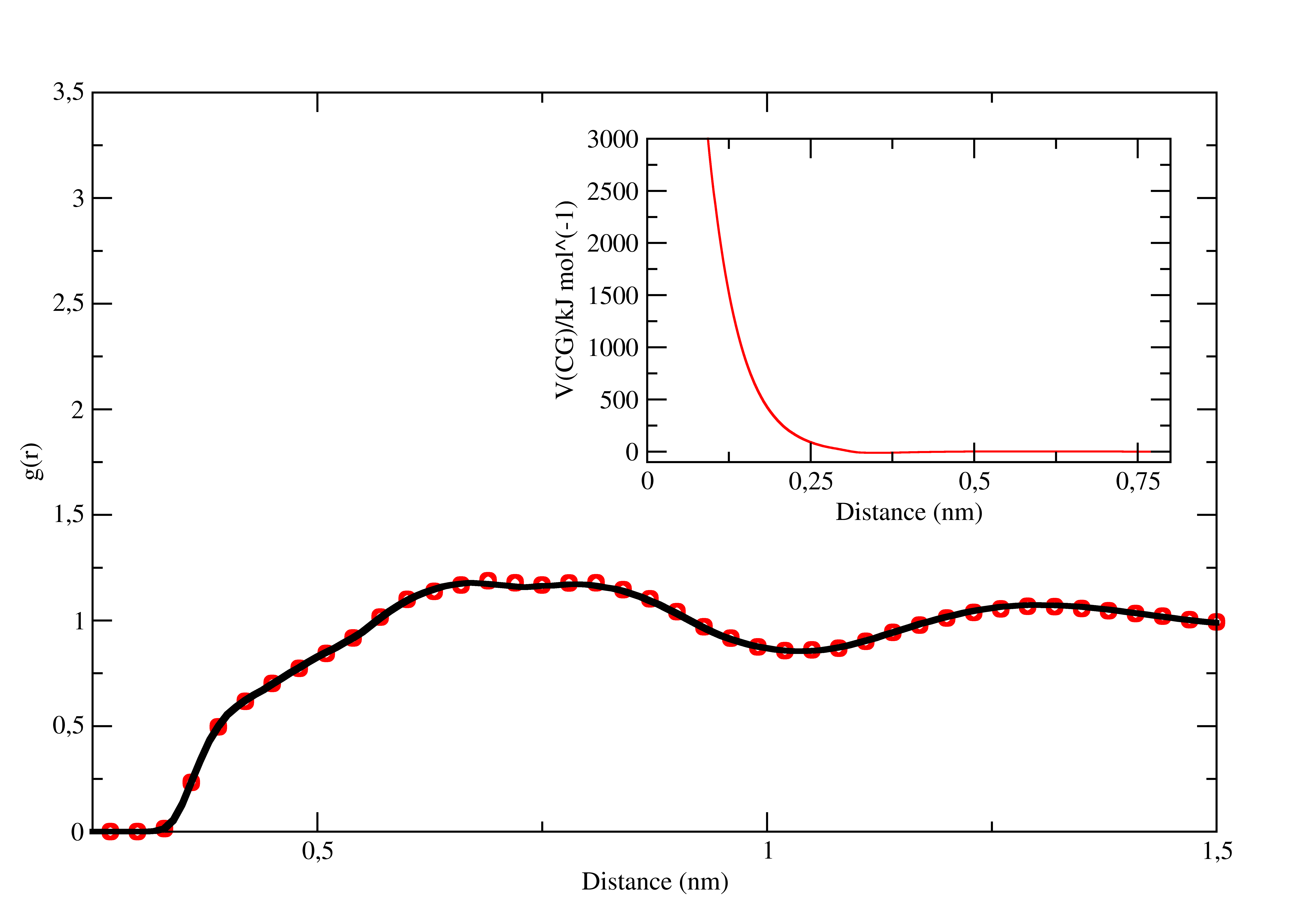}
   \caption{Equivalent of Figure \ref{catcatnc} for the charged coarse-grained model.}
 \label{catcatc}
 \end{figure}

The coarse-grained particles are not charged and once they enter into the hybrid region they slowly start to switch on the electrostatic interactions and thus, effectively, acquire the corresponding charge according to the atomistic model (vice versa ions slowly loose their charges when transit from the atomistic to the coarse-grained region). According to the theoretical principles of GC-AdResS (in which we count also Ref.\cite{gas}), the model used carries more details and thus it should lead already to good results; indeed this is true as it will be shown in section \ref{results}. The reason why a generic liquid (or mixture if we consider each ion a species {\it per se}) was not used is that it is not convenient from the numerical point of view; in fact the reproduction of the radial distribution functions allow the ions to be at a sufficiently large distance when just entering the hybrid region, instead a generic liquid does not always assure that this can happen and thus leads to numerical instabilities. 
In general, in a very simplified spherical coarse-grained model as the one we employ here,  the methyl groups of cations are forces in a spherical region and they may represent a problem when cations acquire atomistic resolution. In fact two methyl groups belonging to two different cations, due to the spherical rotational invariance of the coarse-grained model, may be too close independently from the accuracy of the coarse-grained model. In such case, very small capping forces are applied, as done for large polymers \cite{lupolj}, to avoid this problem. Capping forces could be used also for the case of a generic coarse-grained model, but the use of the IBI is more general and straightforward. The conceptual disadvantage of a neutral coarse-grained model is that the global system, that is the atomistic region and the reservoir, may instantaneously be charged and thus be characterized by possible instantaneous spurious net electrostatic (artificial) interactions. Moreover when ions just enter into the atomistic region the ion-ion electrostatic interaction may still not be optimal and thus the equilibration while crossing the hybrid region to enter in the atomistic region could be less smooth than one may desire. On this purpose we have tried a second option, we derive the coarse-grained potential using the IBI procedure but keeping the corresponding charge in the coarse-grained model. This implies that the potential refined at each step of the IBI iteration is such that its sum with the electrostatic potential (untouched by the IBI procedure) must lead to the reproduction of the radial distribution function of target (see Figs.\ref{ananc}, \ref{ancatc}, \ref{catcatc}). In this way possible limitations of a neutral coarse-grained model reported above are removed. However, we will see in section \ref{results}, that the two models, for the properties considered, are essentially equivalent (with difference of $2 \%$ at worst).

\section{Technical Details}
\label{technical}
All reported simulations were performed with the package GROMACS \cite{gromacs}. The force field parameters for 1,3-dimethyl-imidazolium chloride were taken from Dommert et al. \cite{chemphyschemrev}. 
We set up two systems, the first one (350 ion pairs) was used to derive two coarse grain potentials, which we then transferred to the larger system (1000 ion pairs). 
We optimized both systems using full atomistic NpT calculations. The simulation temperature was set to 400 K, the time step was set as 2 fs and the electrostatic interactions were calculated through particle mesh Ewald technique (PME). For the first 2 ns we used the Berendsen barostat \cite{berendsen}, after that we switched to the Parrinello-Rahman barostat \cite{par_rah} for the following 2 ns. We monitored the box size and considered that we have reached convergence, when the changes of the box length was of the order of 0.0001 nm. The density obtained is 1155 $kg/m^3$ which agrees well with the experimentally found density of 1165 $g/cc$ \cite{jcp_exp}. For 350 ion pairs we obtained a cubic box with 4.08525 nm box length, while for the 1000 ion pairs we got a box with a=12.09408 nm, b=c=4.03137 nm. 
Furthermore, the radial distribution functions were obtained after 10 ns full atomistic NVT simulations at 400K and with 2 fs time steps. The results are shown in Fig. \ref{collgr} and agree well with those from literature \cite{balasub_cpl}. 
The 350 ion pairs configuration was used to derive the two coarse-grained models. We used an inverse Boltzmann Iteration, IBI, procedure \cite{ibi} to reproduce the radial distributions functions of the full atomistic target system (see Figs.\ref{anannc}, \ref{ancatnc}, \ref{catcatnc} and for the charged model see Figs.\ref{ananc}, \ref{ancatc}, \ref{catcatc}).
After we derived those tabulated potentials, we used the configuration from the 1000 ion pairs to set up the GC-AdResS system. An atomistic region of a=3 nm is bordered by hybrid regions of the same length, while the remaining a=3.09408 nm is coarse-graining region. As in standard AdResS simulations, a Langeving thermostat is used with $\Gamma=10 ps^{-1}$. The electrostatic interactions, as usually done in AdResS, were treated by the generalized reaction field method with self-consistent dielectric constant as calculated by Gromacs \cite{junghans}. The use of the charged coarse-grained model implied that the reaction field method is used also in the coarse-graining region where it applies in the same way as in the atomistic region. The thermodynamic force through the iterative procedure was considered converged after 8 iterations, sufficient to reach an accuracy of $3\%$ for the particle density and for the ion-ion radial distribution functions (compared to the full atomistic of reference). The use of the charged coarse-grained model implied that the reaction field method is applied also to the coarse-graining region with the effective ionic strength (dielectric constant) one would have in the reaction field method applied to the full atomistic simulation.

\section{Results}
\label{results}
\begin{figure}
   \centering
   \includegraphics[width=0.75\textwidth]{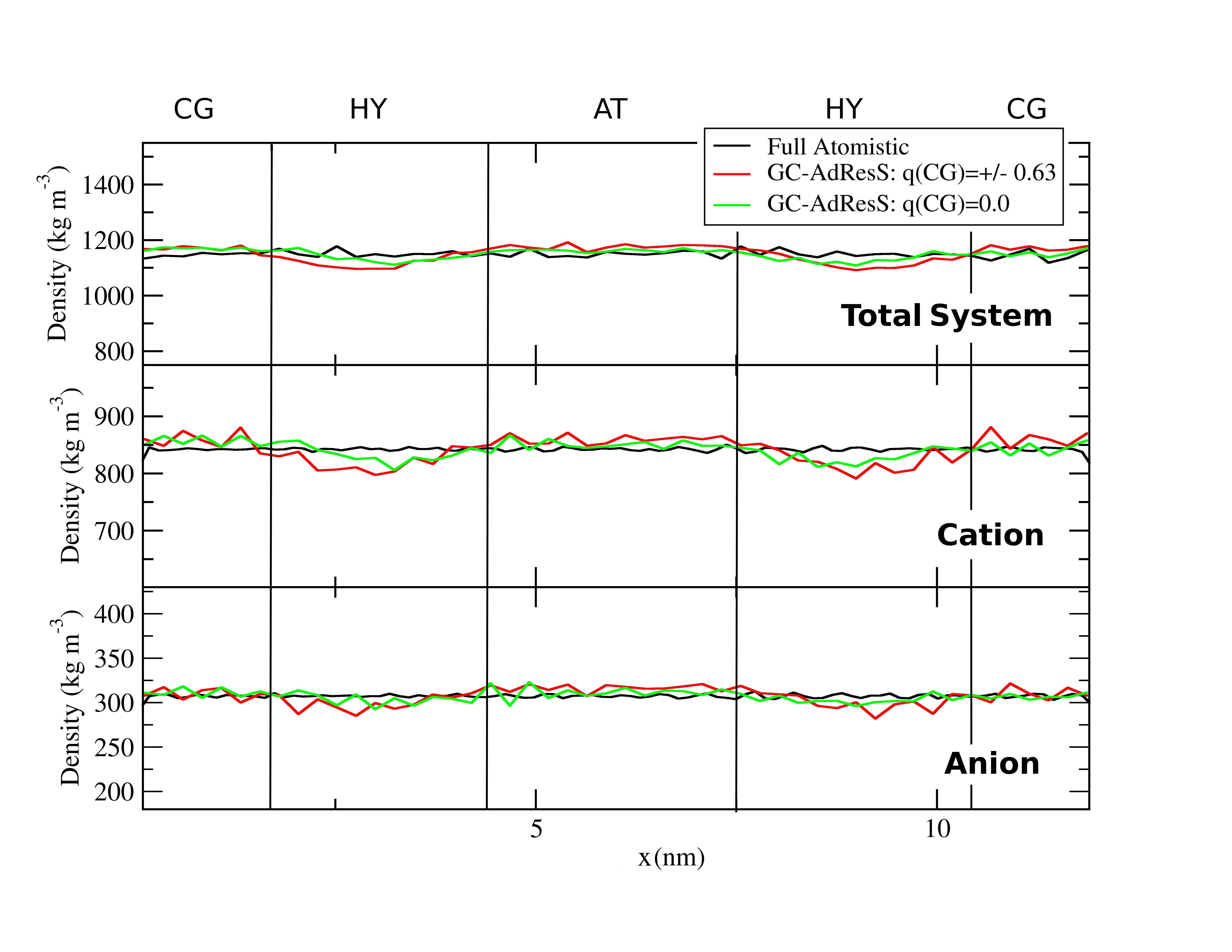}
   \caption{Particle density as a function of the position in space along the direction where the change of resolution occurs. The black line refers to the full atomistic simulation, the red to the GC-AdResS simulation with uncharged coarse-grained model and the red line refers tho the GC-AdResS simulation with charged coarse-grained model. The top panel refers to ion-pairs, the middle panel to cations and the bottom panel to anions. As expected the largest deviation occur in the hybrid region with an upper value of about $7\%$. However in the atomistic region, which is the region of interest the largest deviation is about $3\%$.}
 \label{dens}
 \end{figure}

\begin{figure}
   \centering
   \includegraphics[width=0.75\textwidth]{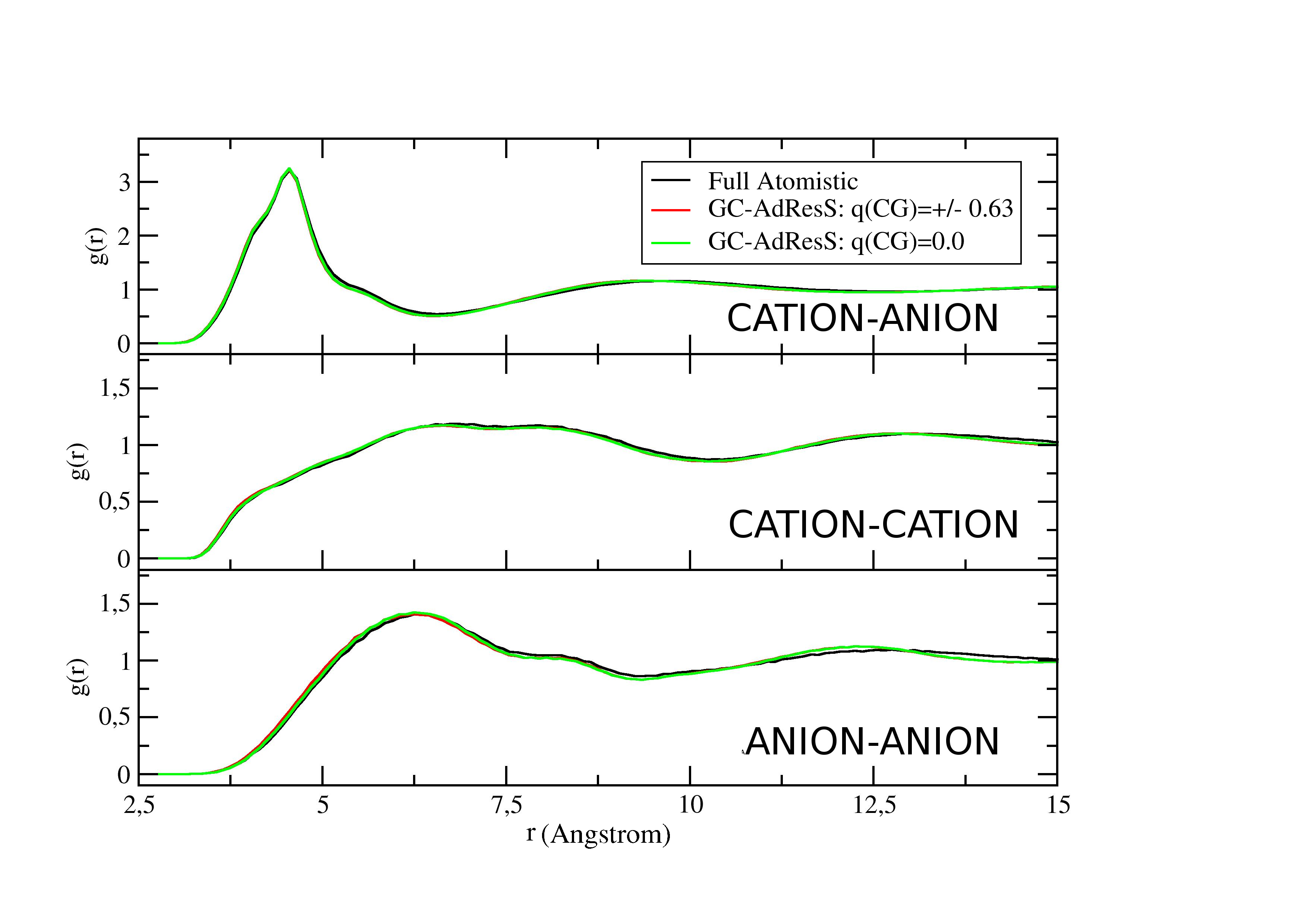}
   \caption{Cation-anion (top), cation-cation (middle) and anion-anion (bottom, radial distribution function calculated in the atomistic region of GC-AdResS for both coarse-grained model and compared with the corresponding quantity calculated in the equivalent region of a full atomistic simulation. The accuracy is highly satisfactory and the curves actually overlap.}
 \label{collgr}
 \end{figure}

\begin{figure}
   \centering
   \includegraphics[width=0.3\textwidth]{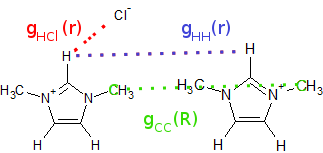}
   \caption{Pictorial illustration which shows the type of atoms that define $g_{HCl}$(r); $g_{CC}$(r); $g_{HH}$(r) respectively.}
 \label{cartoon2}
 \end{figure}

\begin{figure}
   \centering
   \includegraphics[width=0.75\textwidth]{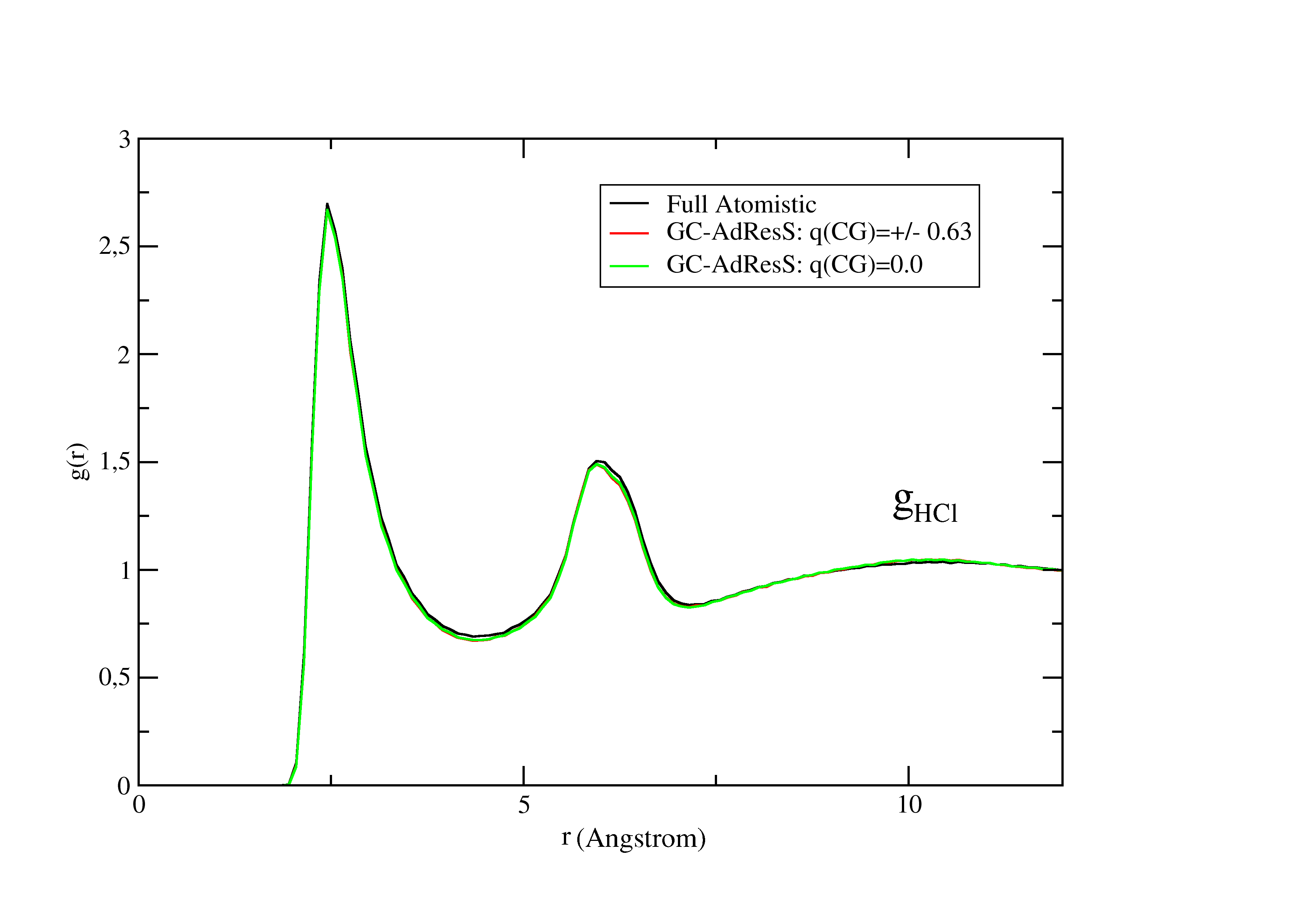}
   \caption{Results for the $g_{HCl}$(r) function. The calculation and the legend is the same as that of Figure \ref{collgr}. Also in this case the agreement is satisfactory.}
 \label{ghl}
 \end{figure}

\begin{figure}
   \centering
   \includegraphics[width=0.75\textwidth]{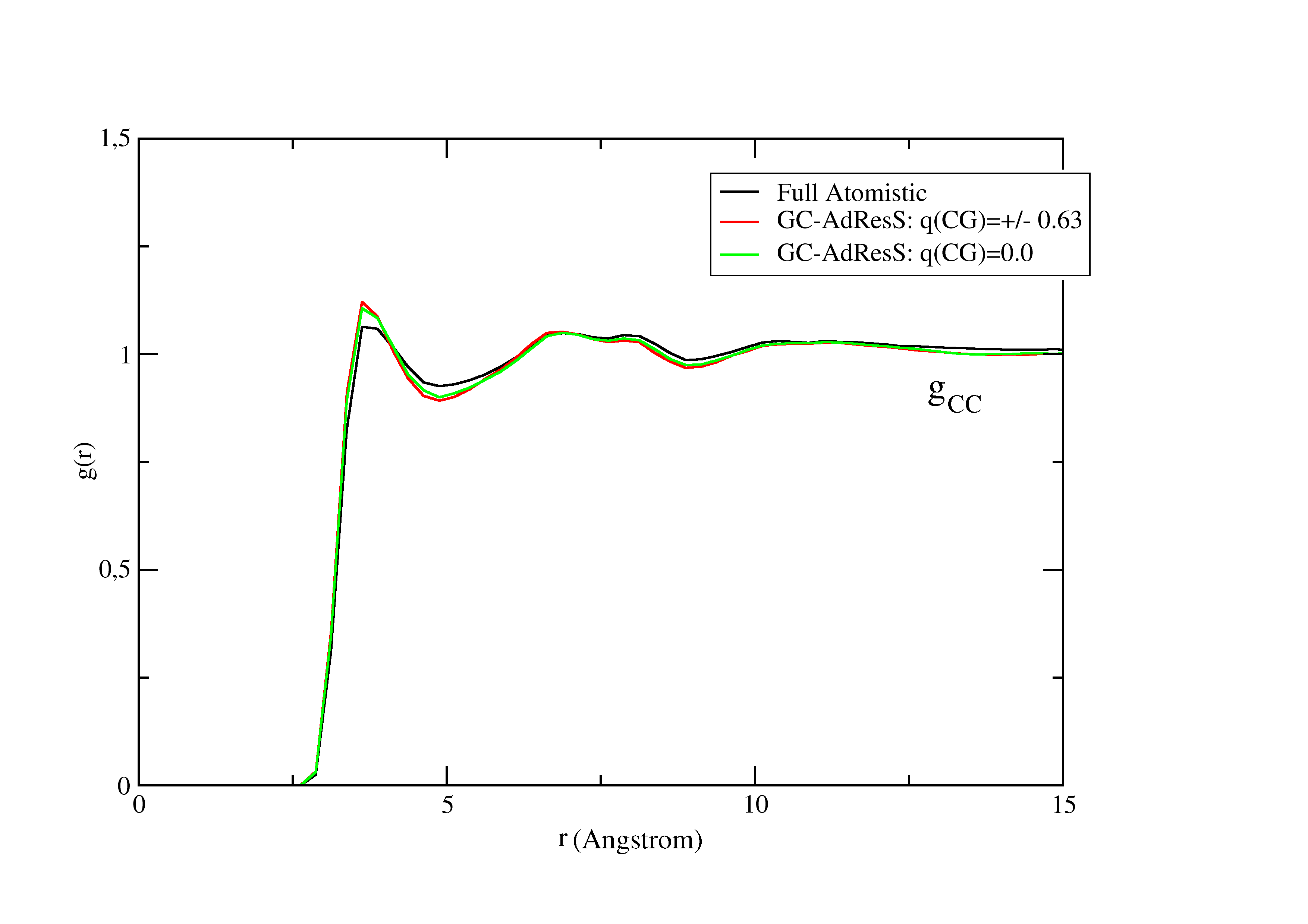}
   \caption{Equivalent of Figure \ref{ghl} for the $g_{CC}$(r) function. As before, the agreement is satisfactory.}
 \label{gcc}
 \end{figure}

\begin{figure}
   \centering
   \includegraphics[width=0.75\textwidth]{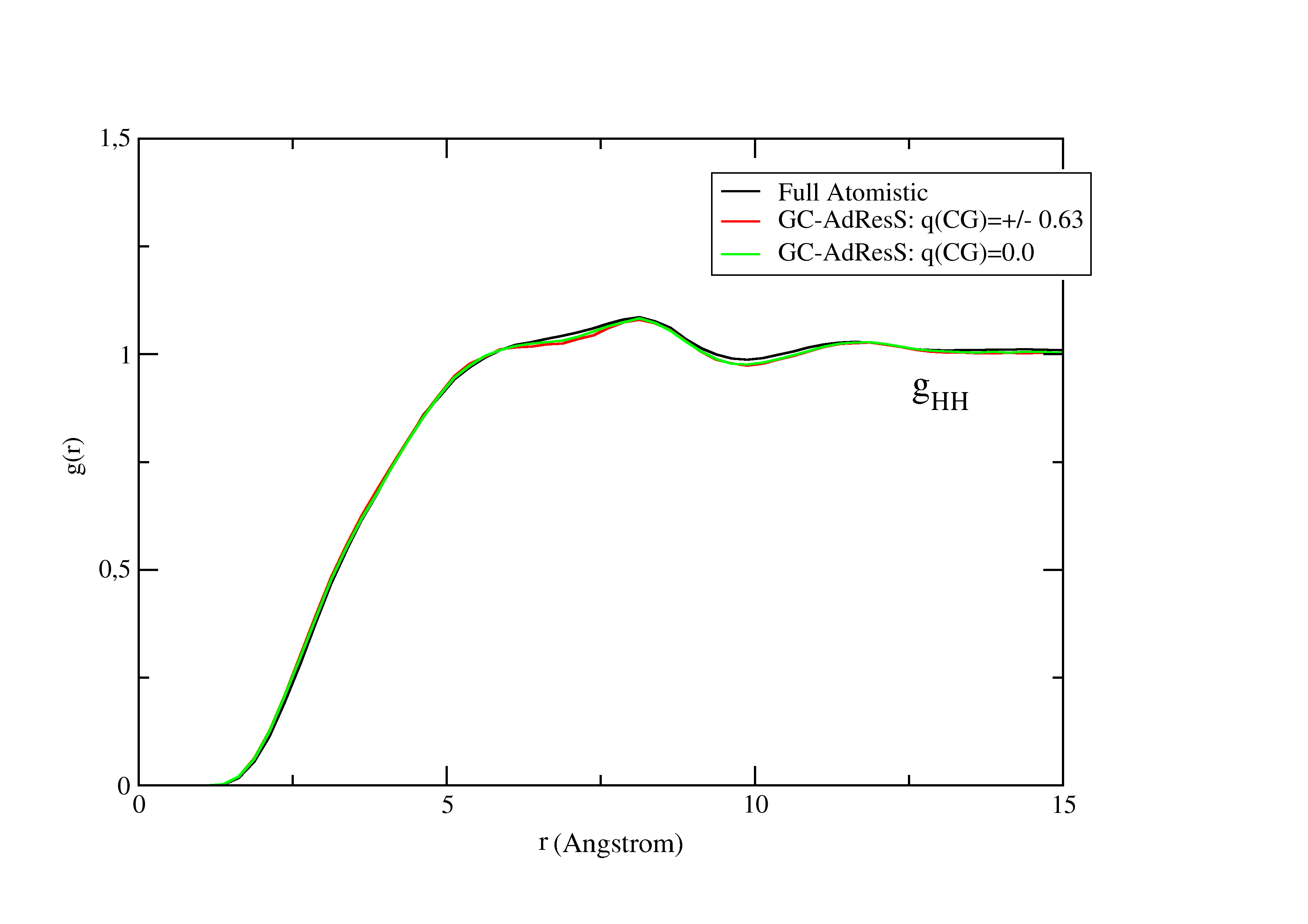}
   \caption{Equivalent of Figure \ref{ghl} for the $g_{HH}$(r) function. As before, the agreement is satisfactory.}
 \label{ghh}
 \end{figure}

The simulation set up of AdResS reported in section \ref{technical} was specifically chosen in order to test the robustness of the method in critical technical conditions, that is we use a relatively small reservoir (hybrid plus coarse-grained region). The method would perform at its best when the reservoir is very large compared to the atomistic region, but if it performs sufficiently well in critical technical conditions, obviously its performance would be satisfactory in ideal conditions. Our aim in this work is to show that basic distribution functions calculated in the atomistic region agree in a satisfactory way with those calculated in the equivalent sub-region of a full atomistic simulation of reference. In fact distribution functions such as particle density and radial distribution functions correspond to the first and second order terms of the statistical distribution in the atomistic region respectively (see \cite{prx}). This implies that if the distribution functions of the atomistic region of GC-AdResS agree in a satisfactory way with the corresponding functions calculated in the equivalent sub-region of the full atomistic simulation of reference, then up to (at least) second order, statistical averages calculated over the atomistic region are equivalent to those calculated in the sub-region of a full atomistic simulation. Moreover, a sub-region in a full atomistic region (of a size that is statistically relevant) is a natural open system embedded in a thermodynamic environment provided by the rest of the system; this implies that if the atomistic region of GC-AdResS reproduces the statistical distribution (at least up to a certain order) of the sub-region of the reference full atomistic system, then the atomistic region is equivalent (up to at least a certain order) to a subsystem in a given thermodynamic bath/reservoir as the open subsystem of the full atomistic simulation. In this perspective, the first quantity to consider is the particle density, in fact if the coupling would induce evident artificial effects, the density will display sizable deviations from the results of the full atomistic simulation. Fig.\ref{dens} reports the comparison between the full atomistic simulation and the GC-AdResS results for the two different coarse-grained models. 
The region of major interest is the atomistic region and the largest deviation from the density of reference of the full atomistic simulation is given by the charged coarse-grained model. The difference is not larger than $3\%$ which, considering also the critical technical conditions, we consider highly satisfactory. Next we considered the cation-cation, cation-anion and the anion-anion radial distribution functions; the results are reported in Fig.\ref{collgr}.
Also in this case, the GC-AdResS results with both coarse-grained model satisfactorily agree with the reference full atomistic simulation and they actually overlap (the error is within the thickness of the lines). However, a deeper check is to show that microscopic radial distribution functions are satisfactorily reproduced, since the atomistic region must be characterized by accuracy at the very atomistic level. For this reason we calculated three representative atom-atom radial distribution function, $g_{HCl}$(r), $g_{CC}$(r), $g_{HH}$(r), (for the definition see Fig.\ref{cartoon2}). Results are reported in Figs.\ref{ghl},\ref{gcc},\ref{ghh}, in all cases the agreement is satisfactory.

\begin{figure}
   \centering
   \includegraphics[width=0.75\textwidth]{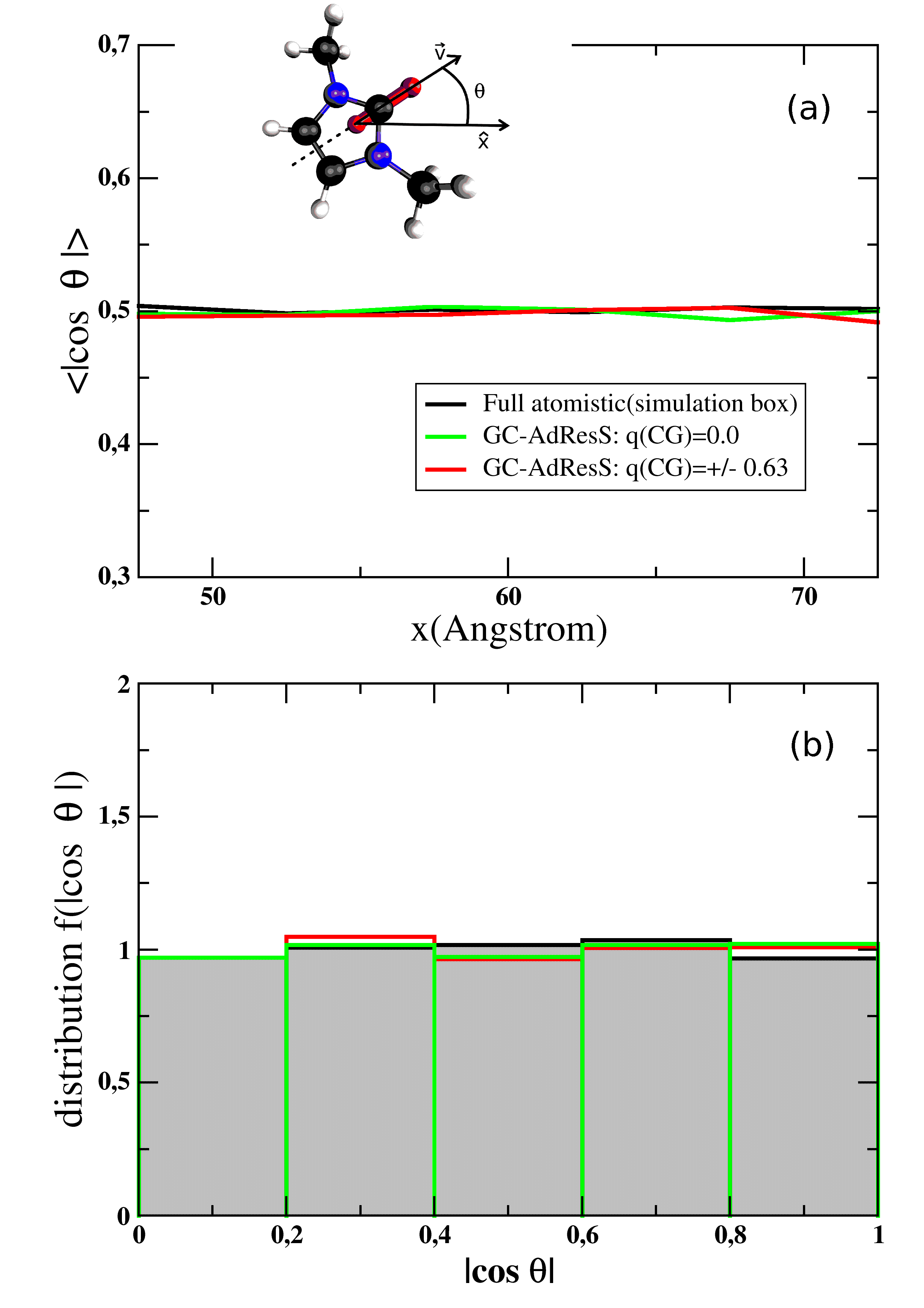}
   \caption{\red{Orientational order parameter: $|\cos\theta|=|\frac{{\bf v}}{{|{\bf v}|}}\cdot \hat{x}|$. Panel (a) shows the ensemble average of $|\cos\theta|$ as a function of the position in the atomistic region. Panel (b) shows the probability distribution of $|\cos\theta|$}.}
 \label{parord}
 \end{figure}
\begin{figure}
   \centering
   \includegraphics[width=0.75\textwidth]{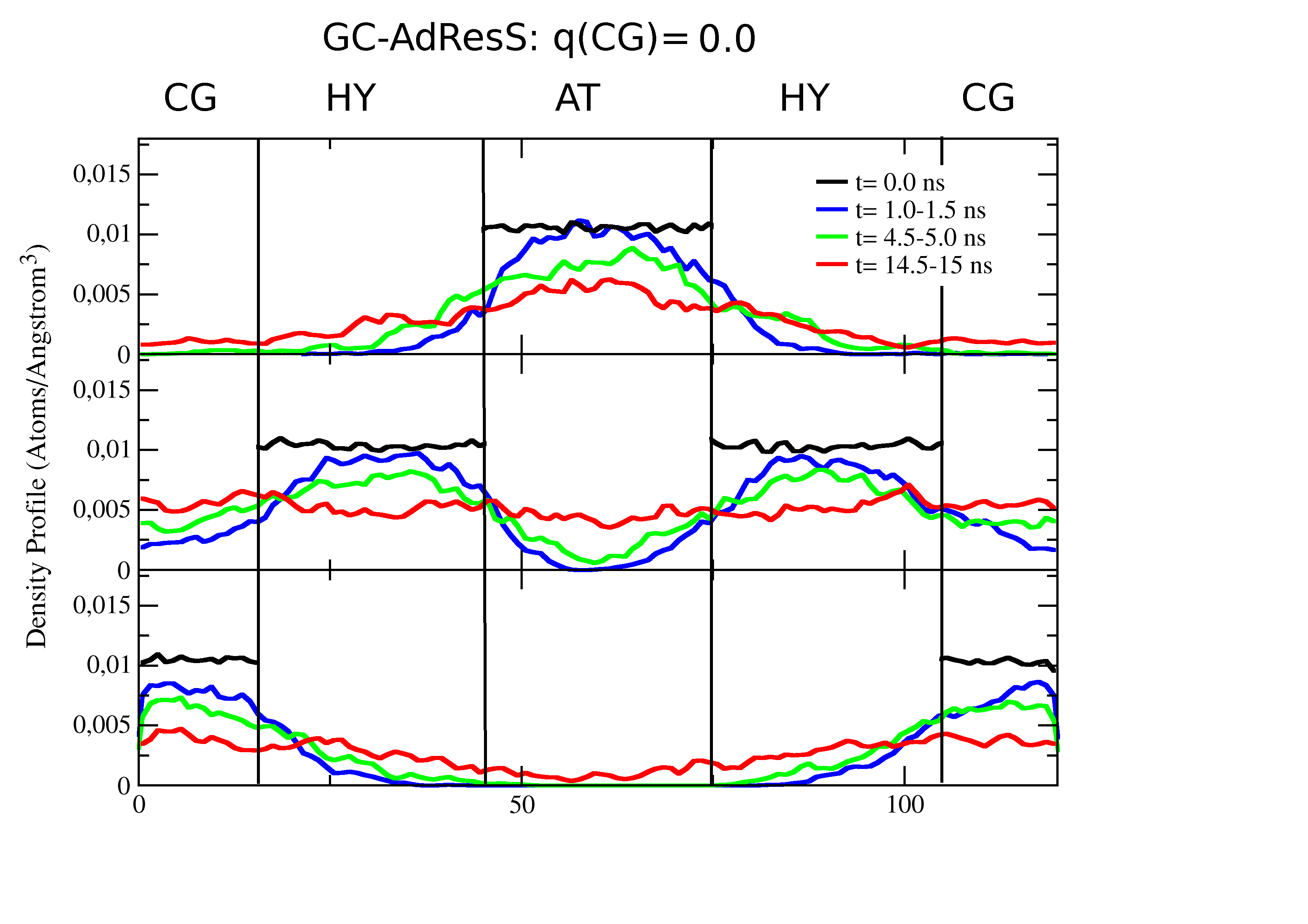}
   \caption{\red{Evolution in time of an instantaneous distribution profile along the trajectory for the ion pairs that are at time, $t=0$, located in the atomistic region (top panel),  hybrid region (middle panel) and in the coarse-grained region (bottom panel). Here we consider a GC-AdResS simulation with uncharged coarse-grained model. The results indicate that there is exchange of ion pairs among different regions and it is consistent with the GC-AdResS set up.}}
 \label{diffnc}
 \end{figure}
\begin{figure}
   \centering
   \includegraphics[width=0.75\textwidth]{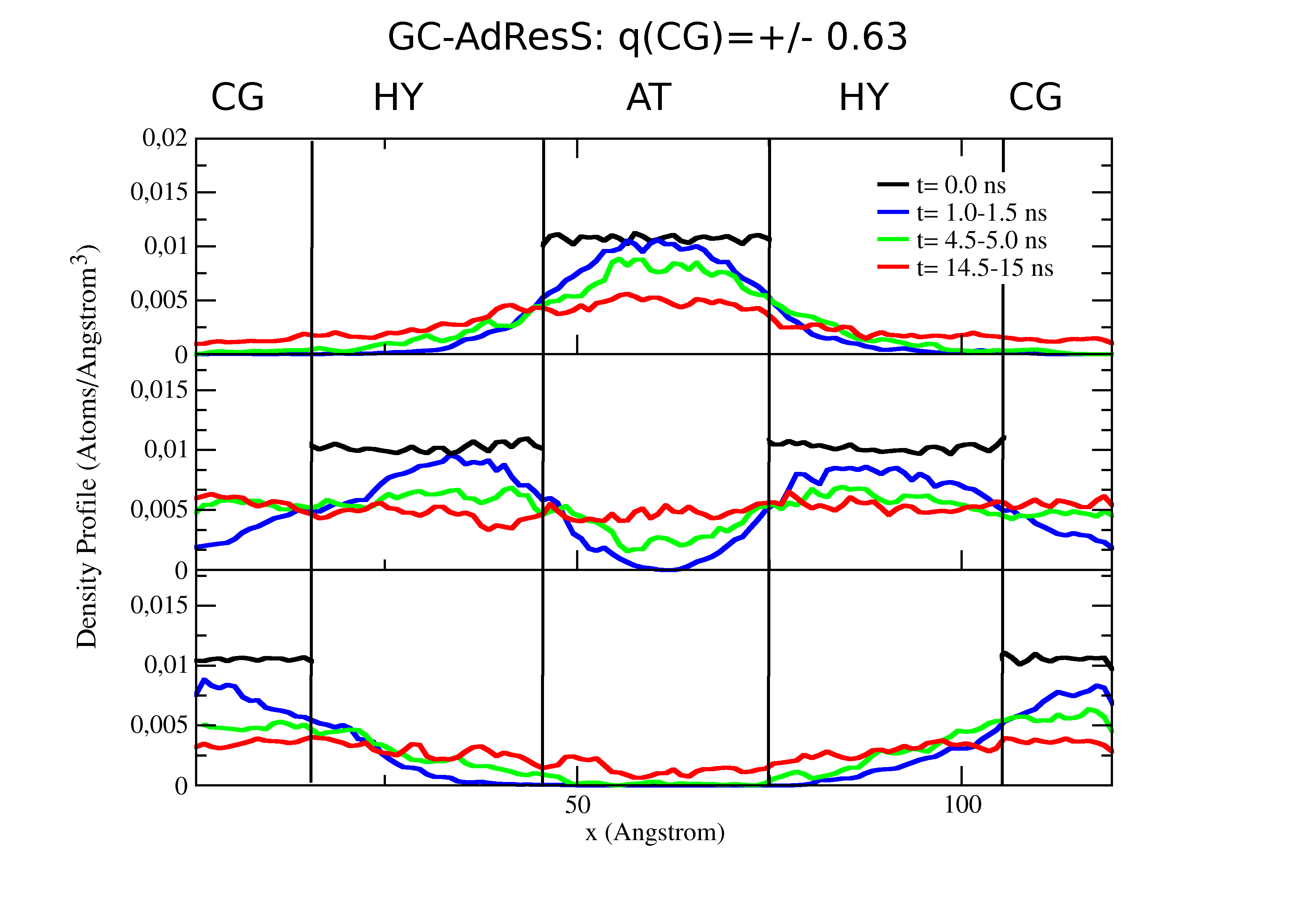}
   \caption{\red{Equivalent of Figure \ref{diffnc} for the charged coarse-grained model. As for the uncharged model, also in this case the flux behaves properly.}}
 \label{diffc}
 \end{figure}

\red{Regarding structural properties, one must check also the possibility of an artificial orientational alignment of the molecules (in this case of the cations) due to the interface introduced by the hybrid region. In order to perform this study we define an orientational order parameter, $|\cos\theta|$, calculated with respect to the $x$-direction (direction of change of molecular representation): $|\cos\theta|=|\frac{{\bf v}}{{|{\bf v}|}}\cdot \hat{x}|$. Here ${\bf v}/|{\bf v}|$ is the unit vector of the principal axis of symmetry of the cation and $\hat{x}$ is the unit vecor along the $x$-direction (see inset of Fig.\ref{parord} for a pictorial representation). Fig.\ref{parord} shows (a) the orientational order parameter as a function of the $x$ position in the atomistic region and (b) its probability distribution; as it can be seen there is in general no preferential orientation of the cations and in particular, even at the interface with the hybrid region the perturbation w.r.t. the full atomistic case, due to the change of resolution, is negligible}. 
According to the procedure usually employed for the validation of the AdResS approach, it is important now to show that there is a proper exchange of particles among the different regions. In fact it must be excluded the possibility that artificial barriers in the hybrid region hinter the exchange of molecules. If molecules do not diffuse from one region to the other, the results obtained for the atomistic region would correspond to those of an effectively artificial closed system. Figs.\ref{diffnc}, \ref{diffc}, show the diffusion of the instantaneous particle distribution from each region to the others and thus a proves that the method behaves reasonably well. \red{The instantaneous exchange is slightly asymmetric because while the ions in the atomistic region diffuse as in a full atomistic simulation, the dynamics in the coarse-grained region is instead faster. In fact a consequence of the reduced number of degrees of freedom is a time scale difference in the dynamics of the coarse-grained system compared to the atomistic one}. This problem is well known for the AdResS simulation and can be fixed by slowing down the dynamics in the coarse-grained and hybrid region with an increase of the friction parameter in the Langevin thermostat. However it has been shown that statistical averages in the atomistic region are not affected by the asymmetry \cite{wat2}.
Moreover a quantity related to the exchange of particles and to the statistical distributions is the particle probability distribution in the atomistic region, $P(N)$, where $N$ is the number of ion pairs. Its shape should follow a Gaussian function and should be compatible as much as possible, within a certain accuracy, with the corresponding quantity calculated in the equivalent sub-region of the atomistic region. 
\begin{figure}
   \centering
   \includegraphics[width=0.75\textwidth]{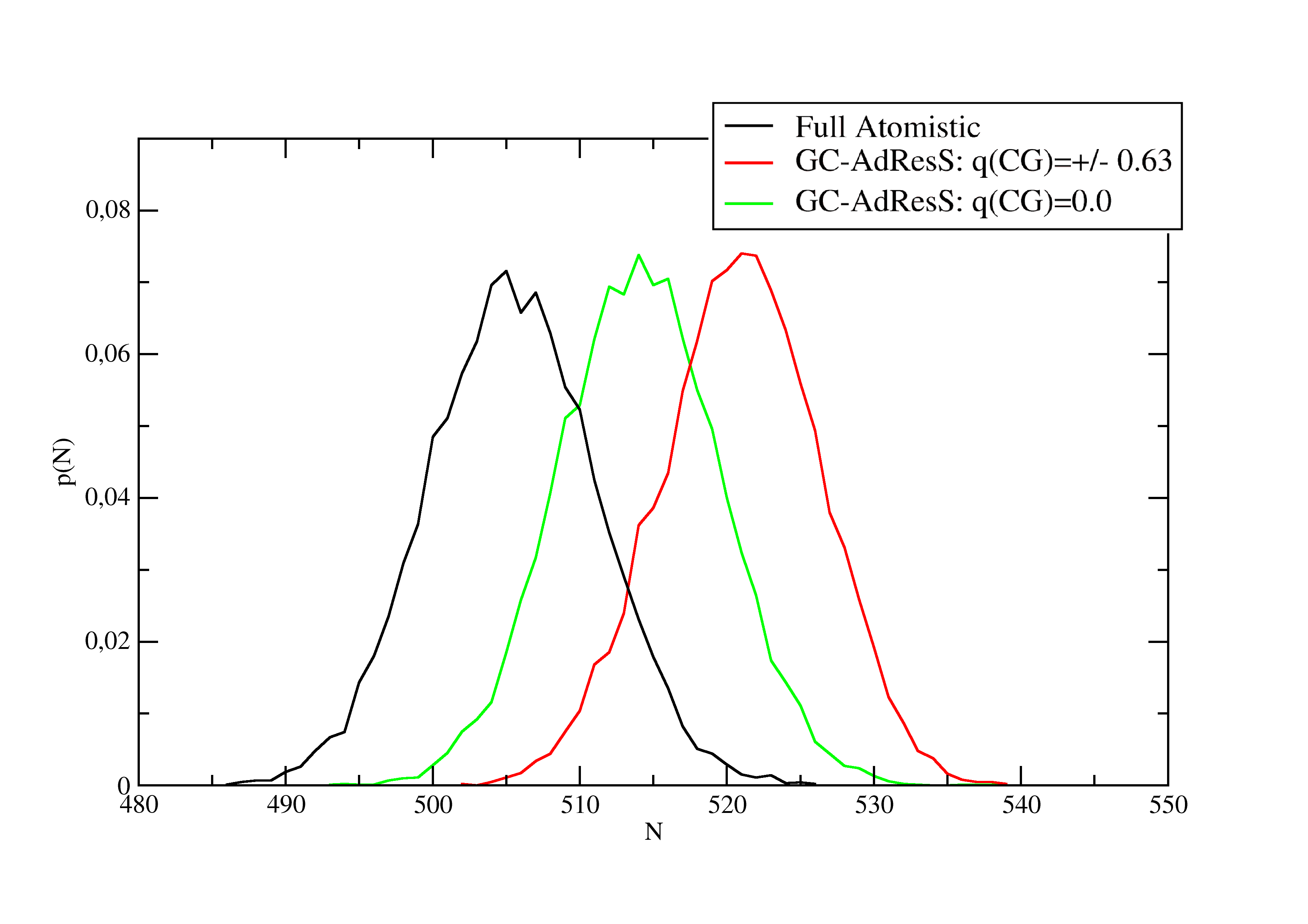}
   \caption{The ion pairs number probability density in the atomistic region compared with the corresponding quantity in the equivalent sub-region of a full atomistic simulation. Compared to the results of the full atomistic simulation, there is a systematic shift of the peak which in the worst case (charged coarse-grained model) is of $3\%$. This is due to the $3\%$ excess of density reported in Figure \ref{dens} which we have chosen as acceptable accuracy.}
 \label{pn1}
 \end{figure}
Fig.\ref{pn1} shows $P(N)$ calculated for GC-AdResS for the two models and compared with the result of the full atomistic simulation. There is a systematic shift of the location of the peak (number of ions with the highest probability), the largest deviation occurs for the case in which the charged coarse-grained model is found. This deviation of the most probable number of molecules is about $3\%$ and it is due to the systematic $3\%$ excess of particle density. 
\begin{figure}
   \centering
   \includegraphics[width=0.7\textwidth]{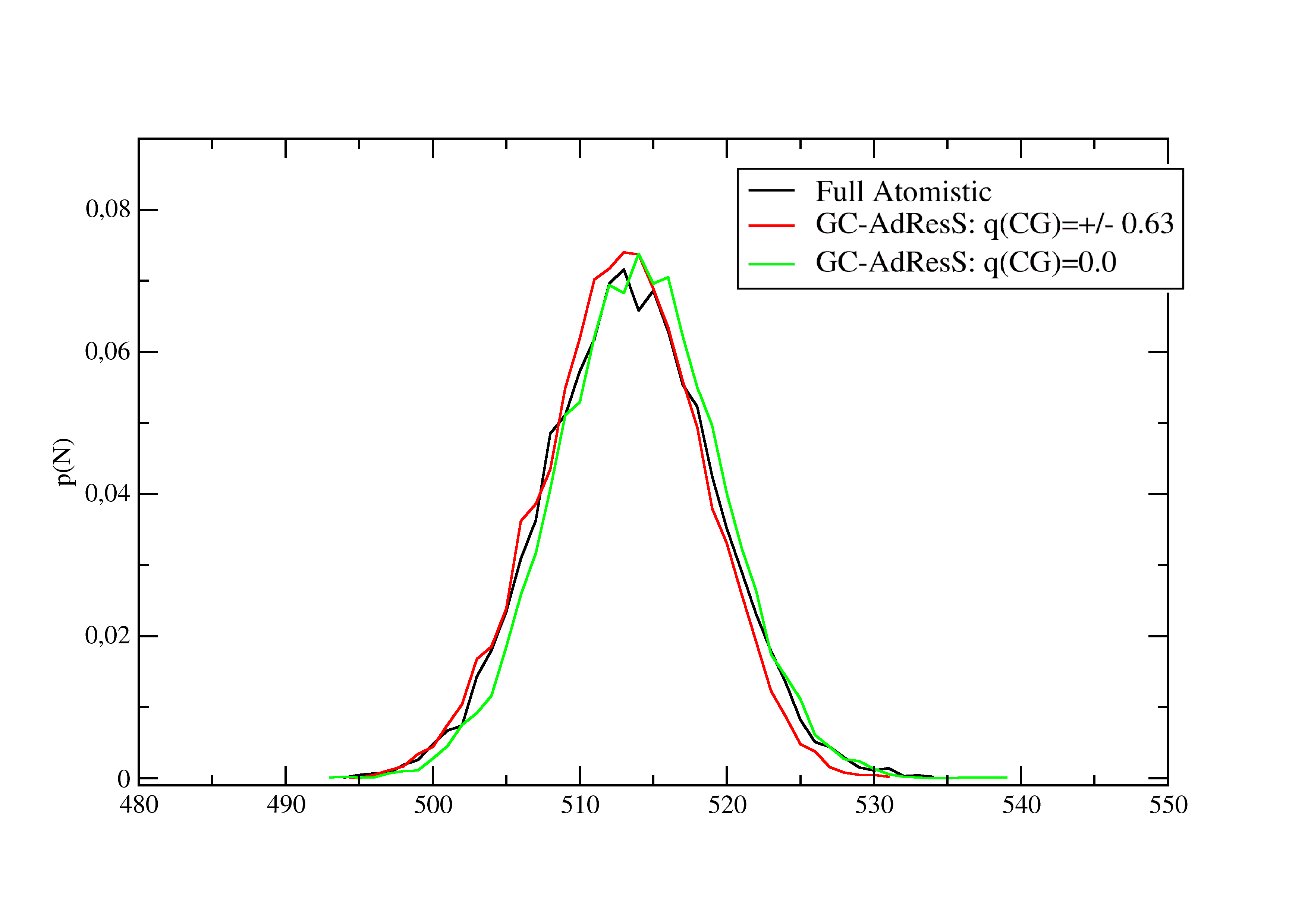}
   \caption{The three curves of Figure \ref{pn1} shifted in such a way that they are centered at the same point. The result shows that the shape of the three curves is very similar and thus indicates a proper statistical behavior of the GC-AdResS simulation.}
 \label{pn2}
 \end{figure}
However, Fig.\ref{pn2} shows the three curves when they are superimposed (with a systematic shift along the axis of $N$), the curves show a satisfactory agreement regarding their shapes. This latter is actually the relevant aspect of the problem because a systematic shift of 15 ion pairs over 500 ion pairs is numerically negligible while a sizable divergence of the shape of the distributions would imply a non valid statistical behavior of the GC-AdResS systems. Once again it should be underlined that we are treating critical technical conditions and thus the results here can be considered as a sort of upper bound of the disagreement for studies with ideal technical conditions. Moreover, one may run further iteration steps for the thermodynamic force and reach an agreement for the particle density within $1-2\% $ of accuracy.

\section{Conclusion}
We have shown the applicability of the GC-AdResS to the study of 1,3-dimethylimidazolium chloride which represents the prototype of a large class of IL's.
We have performed our study in technical conditions which can be considered a ``worst case scenario'' of a GC-AdResS set up and have shown that the results are satisfactory. This study is relevant for two main reasons: (a) GC-AdResS can be applied as a tool of analysis in order to identify the essential atomistic degrees of freedom that characterize a given property of the system; it has been proven very useful for other systems and thus it is, in our view, an important tool for IL's as well, (b) AdResS has already allowed the coupling of particle-based regions  with the continuum; this was done for generic fluids and liquid water but one may think to make this extension for IL's as well, so that one can have systems with virtually infinite size and span a large spectrum of scales in a concurrent fashion. In this perspective, this work represents the initial step for an open boundary molecular dynamics simulation of IL's \cite{matejepjst}. \red{Future applications of the adaptive method to ionic liquids in the short term concern the study of other ionic liquids. In particular those where the extension of the current model can be done straightforwardly, e.g. those systems where the molecular structural modification of the cations involves the addition of alkyl groups of a certain size, a molecular structural aspect that recent technical advancements of AdResS can now systematically treat \cite{lupolj}. Next, GC-AdResS allows to treat an atomistic spherical region embedded in a large reservoir of structureless molecules, thus the dependence of structural and dynamic properties as a function of the size of the atomistic region can be systematically studied. By comparing the results of different simulations, each characterized by an atomistic region of a different size, one can determine the connection between time and length scales of given properties. For example one can calculate the {\it hydrogen bond-hydrogen bond} autocorrelation function as a function of size of the atomistic region. Such a quantity links the information concerning the relevance of the hydrogen bonding network of the bulk for the formation of hydrogen bonds at local level, i.e. in the atomistic region (length scale), to the life time of a hydrogen bond (time scale) (see also Refs.\cite{njp,lucpc}, where this study was done for liquid water). Moreover the comparison between the results obtained with different ionic liquids will clarify the effect of the specific molecular chemical structure of the ions on the scales interplay.  
Such a study would allow for a direct check and a deeper understanding of the hypothesis of rattling ions in long-living ion cages proposed in literature \cite{hypo1,hypo2,hypo3,hypo4,hypo5}.}

\section*{Acknowledgment}
This research has been funded by Deutsche Forschungsgemeinschaft (DFG) through grants CRC 1114 (project C01) LDS and by the European Community through project E-CAM for CK and LDS.

\end{document}